\documentstyle[epsfig,psfig,aps,prl,multicol]{revtex}
\begin{document}
\draft
\title
{
Competing phases in the high field phase diagram of (TMTSF)$_2$ClO$_4$
}
\author
{
S. Haddad$^1$, S. Charfi-Kaddour$^1$, C. Nickel$^2$, M. H\'eritier$^2$ and R. Bennaceur$^1$
} 
\address{
$^1$ Laboratoire de Physique de la Mati\`ere Condens\'ee, D\'epartement de Physique,\newline
Facult\'e des Sciences de Tunis, Campus universitaire 1060 Tunis, Tunisia \newline
$^2$ Laboratoire de Physique des Solides (associ\'e au CNRS), Universit\'e de Paris-Sud 91405 Orsay, France \newline
}
\maketitle
%
%
\begin{abstract}
A model is presented for the high field phase diagram of (TMTSF)$_2$ClO$_4$, 
taking into account the anion ordering, which splits the Fermi surface in two bands.
For strong enough field, the largest metal-SDW critical temperature corresponds to the N=0
phase, which originates from two intraband nesting processes.
At lower temperature, the competition between these processes
puts at disadvantage the N=0 phase vs. the N=1 phase, which is due to interband
nesting. A first order transition takes then place from the
N=0 to N=1 phase. We ascribe to this effect the experimentally observed phase diagrams.
\end{abstract}

\begin{multicols}{2}
The Bechgaard salts, (TMTSF)$_2$X, (X= PF$_6$, ClO$_4$, ReO$_4$...)
exhibit a rich variety of original properties\cite{revue}. One of the most
interesting phenomena is certainly the quantum cascade of Spin Density Wave
(SDW) phases induced by a magnetic field. The simplest case is given by the X=PF$_6$ salt.
In the low temperature metallic phase of this salt, a moderate magnetic
field applied in a direction perpendicular to the most conducting planes
induces a cascade of transitions to SDW phases, exhibiting the quantized
Hall resistance $\rho_{xy}=h/2Ne^{2}$ in the sequence $N=...5, 4, 3, 2, 1, 0$
as the field is increased. The experimental data are well
explained by the Quantized Nesting Model (QNM)\cite{Gor,Heritier,Chaikin,Poil,Maki}: 
in a Fermi liquid approach of the metallic phase,
described by two slightly warped parallel sheets Fermi surface, the orbital
effect of the magnetic field destabilizes the metal by inducing a sort of
Peierls instability to a SDW phase. As the field is varied, the SDW wave
vector adjusts itself to ensure the Peierls condition that the Fermi level
lies in the middle of one of the SDW Landau gaps.

While the experiments in the PF$_6$ salt are well explained, even with a
quite simple Fermi surface model, the case of the ClO$_4$ salt
exhibits distinct deviations from the PF$_6$ behavior
which, definitely, cannot be understood within the QNM
\cite{Naughton,Kang,Mc,Scheven,Uji,Moser,Chung}.
Although the phenomenon of quantized cascade of SDW phases still exists, the
phase diagram is more complex.
In the ClO$_4$ salt, the quantized cascade is observed up to 8 T. When the magnetic
field $H$ is further increased, the metal-SDW second order transition
temperature continues to increase smoothly, but saturates at about 5.5 K for 
$H$ larger than 18 T. In contrast with the behavior of the PF$_6$
salt, a second phase transition occurs inside the domain of stability of the
field induced SDW. This is a first order transition, with a maximum
transition temperature of 3.5 K for $H=20$ T. When $H$ is increased
above 20 T, this transition temperature strongly decreases and vanishes at
28 T. The existence of this first order line for 20 T$< H <$28 T has been 
first reported by McKernan {\it et al.}\cite{Mc}. J. Moser\cite{Moser} 
has proposed that this line should be prolonged for 17 T$< H <$20 T
with a transition temperature strongly decreasing as $H$ decreases
from  20 T to 17 T, which has been confirmed recently by 
Chung {\it et al.}\cite{Chung}. 
The labeling of the field induced phases by a definite 
quantum number N is by no means obvious in this high field
part of the phase diagram. 
Below $0.3$ K, a quantized Hall plateau,
corresponding to N=1 phase has been observed for 8 T$< H <$28 T.
The first order transition at $H=28$ T and low temperature
leads to a highly resistive state, with a non quantized Hall resistance,
which is strongly reminiscent of a N=0 SDW phase\cite{Naughton,Mc}.
These two facts would seem to imply the values of
the quantum number N in the $(H,T)$ plane, since 
N cannot change unless a first order transition line is crossed.
These experimental evidences are at variance with several previous theoretical works,
predicting a "metallic reentrance" as a consequence of the QNM.

Many authors\cite{Braz,Osada,Gor2} have proposed to ascribe the different
behavior of the ClO$_4$ salt to the existence of an ordering of the
perchlorate anions, which occurs at 24 K.
This dimerizes the system along the {\bf b} direction and opens a gap in 
the original two-sheet Fermi surface, giving rise to four open sheets of 
Fermi surfaces.

In this paper, we propose a model to describe properly the 
behavior of the ClO$_4$ salt.
Although the starting point of this work is a generalization of the QNM,
new ingredients are introduced which are substantially important in the determination
of the phase diagram. In particular, two main ideas, completely neglected so far,
turn out to be decisive: (i) the renormalization of the electron-electron scattering strengths
by the low dimensional fluctuations is different for different SDW phases.
(ii) the competition between a usual single wave vector SDW phase and 
an original SDW phase characterized by two coupled order parameters,
which do not compete but on the contrary cooperate to stabilize this phase.
These ideas have not been already discussed in the literature.
The resulting features deduced from our model are consistent with the 
experimental observations listed above, but quite different from that of 
previous works\cite{Braz,Osada,Gor2}.

As in the standard QNM, we approximate the quasi-1D spectrum of Bechgaard
salts, when no anion ordering is present, by a two harmonics dispersion
relation:
\begin{eqnarray*}
\epsilon\left(\vec{k}\right)=v_F\left(|k_x|-k_F\right)-2t_1\cos k_yb -2t_2\cos2k_yb
\end{eqnarray*}
where $k_{x}$ and $k_{y}$ are the electron momenta along and across the
chains, $v_{F}$ is the Fermi velocity and $b$ is the interchain distance.
$t_1$ denotes the effective interchain transfer integrals to nearest neighbors, and the $t_{2}$
term  accounts for the deviation from perfect nesting of the quasi-1D Fermi
surface.

Under a magnetic field applied in the {\bf c} direction, {\bf H}$=\left(
0,0,H\right)$ and in the Landau gauge {\bf A}$=\left(0,Hx,0\right)$,
the non interacting Hamiltonian in the absence of anion ordering takes the
form 
\begin{eqnarray*}
H_{eff}^{0} &=&v_{F}\left[ \left| -i\frac{\partial }{\partial x}\right|
-k_{F}\right] -2t_{1}\cos \left[ -ib\frac{\partial }{\partial y}+ebHx\right]
\\
&&-2t_{2}\cos \left[ -2ib\frac{\partial }{\partial y}+2ebHx\right]
\end{eqnarray*}
In the presence of the anion ordering, which introduces a periodic potential 
$V\left( y\right)=V\cos \frac{\pi }{b}y$, the Brillouin zone is halved
in the $k_{y}$ axis. In the $\left\{ \mid \psi_{k_{x},l}\rangle \right\}$
basis, where $\mid \psi_{k_{x},l}\rangle$ are the eigenstates of 
$H_{eff}^{0}$, the matrix elements of the anion potential $V\left( y\right)$,
considered as a perturbation, are given by (eq.2 of Ref.\onlinecite{Yakov})
\begin{eqnarray}
\left\langle \psi_{\frac{2\pi n}{L},l}\mid V\mid \psi_{\frac {2\pi
n^{\prime }}{L},l^{\prime }}\right\rangle &= &V \left( -1\right)
^{l}J_{l^{\prime }-l}\left( \frac{4t_{1}}{v_{F}G}\right)\nonumber\\ 
&\times &\delta \left( n-n^{\prime }+\frac{LG\left( l-l^{\prime}\right)}{2\pi }\right)
\end{eqnarray}
$L$ is the length of the sample along the chain direction, the integer $l$ is 
used instead of $k_{y}$\cite{Yakov}, $J$ is the Bessel function and $G=ebH$.
Within a perturbative treatment to first order on V, the
diagonal term with $l=l^{\prime }$ in eq.1 splits the energy spectrum of the
total Hamiltonian into two subbands $E^{A}$ and $E^{B}$ given by (see Fig.1):
$
E_{\vec{k}}^{m}=v_{F}\left( \left| k_{x}\right|-k_{F}^{m}\right)
$,
with $m=A,B$ and $k_{F}^{A}=k_{F}-\frac{\Delta }{v_{F}}$ and 
$k_{F}^{B}=k_{F}+\frac{\Delta }{v_{F}}$,
where $\Delta =VJ_{0}\left( \frac{4t_{1}}{v_{F}G}\right)$.

Considering $VJ_{0}\left(\frac{4t_{1}}{v_{F}G}\right)$ as a perturbation is justified 
by an intensive experimental study \cite{Pouget96,Pouget00},
which yields a value of $V\sim T_{AO}/2 \; \sim $12 K.
However, some magnetoresistance (MR) measurements \cite{Yoshino97} have led to
somewhat larger estimates for $V$, but in fairly broad range, because of
a lack of consistent interpretation of MR properties: some of them are
of the order of $t_1$ \cite{Yoshino97}, others are $\sim $50 K \cite{Uji2}.
Therefore, $V$ cannot be estimated without ambiguities and 
difficulties \cite{Lebed2,Osada2}.
It should be stressed that recent theoretical studies suppose that $V$ may be
large ($V \sim t_1$) \cite{Hasegawa_al}. Nevertheless, such values, because
of a too large departure from perfect nesting, cannot account neither for 
the FISDW cascade, nor for the quantum Hall effect observed in (TMTSF)$_2$ClO$_4$.
Indeed, if $V \sim t_1$, the quantum Hall numbers have to be quite large 
($N \sim 50 $ for $H \sim $10 T), in disagreement
with the experimental data.
%
%
\begin{figure}[h]
\epsfxsize=7cm
\epsfysize=3cm
\centerline{\epsfbox{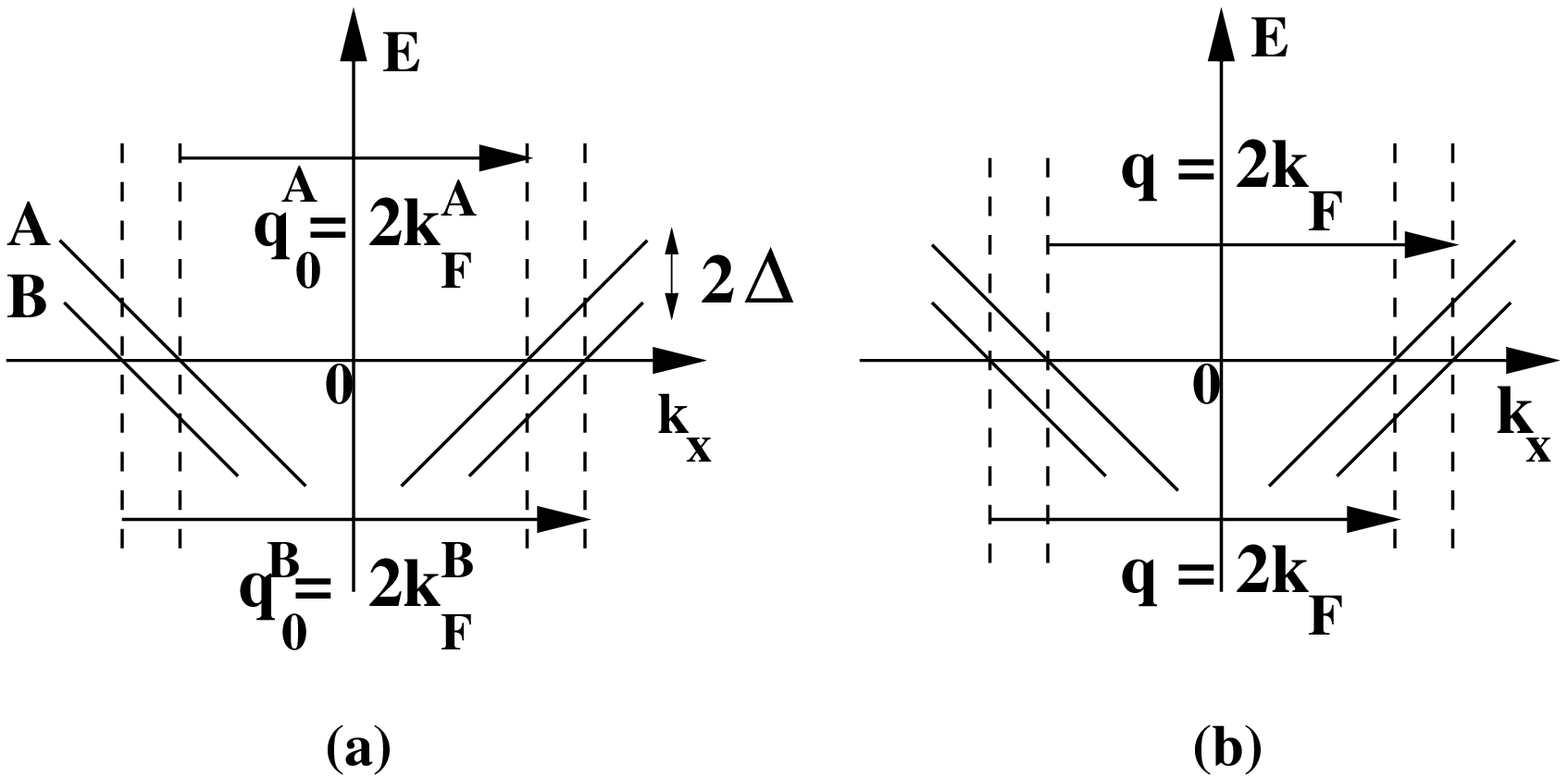}}
{\noindent Fig.1: {\small Band structure of quasi-1D electron system in presence of anion
ordering. $q_{0}^{A}$ and $q_{0}^{B}$ are intraband nesting vectors 
respectively in the A band and the B band while $q$ corresponds to the interband nesting
vector.}}
\end{figure}
The four sheets of the split Fermi surface determine, now,
four different nesting processes : the
intraband nesting A-A or B-B, with two different longitudinal components of
the nesting vector $q_{A}^{0}=2k_{F}^{A}$ and $q_{B}^{0}=2k_{F}^{B}$, 
but also the interband nesting processes A-B and B-A, with only one longitudinal
component of the nesting wave vector $q=2k_{F}$, as in the absence of
anion ordering. In a Fermi liquid approach, the instability of the metallic
phase is discussed by writing the Stoner criterion, which involves the non
interacting spin susceptibilities. It has been shown that the formation of a
SDW phase with an {\it even} ({\it odd}) value of the quantum number is associated with
the  divergence of the {\it intraband} ({\it interband}) susceptibilities \cite{Osada}. 
Since we are interested in the strong
field part of the phase diagram, we restrict the discussion to the phases
corresponding to the two smallest quantum numbers, N=0 and N=1.

The interacting part of the Hamiltonian is given by:
\begin{eqnarray}
H_{int}=\frac{g_2}{2}\sum_{m_i,\sigma}
&\int & d^2\vec{r} \; \;\Psi\left(\vec{r}\right)^{\dagger}_{m_4,-,-\sigma}
\Psi^{\dagger}_{m_3,+,\sigma}\left(\vec{r}\right)\nonumber \\
&\times &\Psi_{m_2,+,\sigma}\left(\vec{r}\right)
\Psi_{m_1,-,-\sigma}\left(\vec{r}\right)
\end{eqnarray}
where $\Psi_{m_i,p,\sigma}$ denotes a fermionic operator for right $(p=+)$
and left $(p=-)$ moving particles. The band label $m_i\; (i$=1-4) corresponds 
to A or B band and $\sigma $ is the spin index.
This Hamiltonian contains only the forward scattering term ($g_2$). The effects
of umklapp scattering ($g_3$) and backward scattering ($g_1$) do not play a central 
role in our model and will be discussed in a forthcoming paper.\newline  
For the N=0 phase, we should define two order parameters
since the nesting vectors of the A and the B bands are different. Let's denote by   
$\Delta^A_0$ and $\Delta^B_0$ the order parameters respectively for the A 
band and the B band which are given by:
$
\Delta^A_0= -\left\langle \Psi^+_{A2\uparrow}
\Psi_{A1\downarrow }\right\rangle \exp ^{iq_{0}^{A}x}$
and 
$\Delta^B_0=-\left\langle \Psi^{+}_{B2\uparrow}
\Psi_{B1\downarrow }\right\rangle \exp ^{iq_{0}^{B}x}
$. We argue for the participation of the two order parameters in the stability of the 
N=0 phase. In fact, the possibility that only one band,
A or B, is gapped at the metal-N=0 transition is not favorable to the formation
of a SDW phase, since the nesting process would only involve one half of the
Fermi surface density of states, which would exponentially reduce the
critical temperature compared to a process involving both the A and the B
bands. Involving these two bands implies the coexistence of two different
SDW's wave vectors.
To lowest order, each SDW has a majority component in its own band and a much smaller
minority component in the other band. The two different SDW's weakly
interact through this overlap of the order parameters in the same band.
At the transition temperature $T_{0}$ from the metallic state to N=0 phase
both pairs of the Fermi surface become simultaneously gapped. \newline
The case of the N=1 phase is simpler since the stability criterion only
includes interband nesting processes. In this case, the N=1 nesting
involves a single wave vector $q_{1}=2k_{F}+G$ , which induces the
formation of a gap at the Fermi level on the four sheets of the Fermi
surface. The order parameter of the N=1 phase is then given by
$\Delta _{1}= -\left\langle \Psi _{A2 \uparrow} ^{\dagger}
\Psi _{B1\downarrow }\right\rangle \exp ^{iq_1x}$.

Based on a microscopic study, we obtain the Landau expansion of the free
energy $F_T$ of the system compared to that of the normal state 
$F_{norm}$:
\begin{eqnarray*}
&F_{T}&-F_{norm}=\frac{a_{0}}{2}\left( \Delta_{0}^{A}\right)^{2}+
\frac{a_{0}}{2}\left( \Delta_{0}^B\right)^{2}+\frac{b_{0}}{2}
\left( \Delta_{0}^{A}\right)^{4}\nonumber \\
&+&\frac{b_{0}}{2}\left( \Delta_{0}^{B}\right)
^{4}+c_{0}\left( \Delta _{0}^{A}\right)^{2}\left( \Delta_{0}^{B}\right)^{2}
+a_1\left(\Delta_1\right)^{2}+b_1\left(\Delta_1\right)^{4}\nonumber \\
&+&\frac {d_{01}}{2}\left(\Delta_1\right)^{2}\left[\left( \Delta_{0}^{A}\right)^{2}
+ \left( \Delta_{0}^B\right)^{2} \right]
\end{eqnarray*}
We find that the coupling term $d_{01}$, which is positive and fairly strong,
leads to an increase of $F_{T}$. When minimizing $F_{T}$ with respect to
$\Delta_{0}$ and $\Delta_{1}$ ($\Delta_{0}\equiv \Delta_{0}^{A}=\Delta_{0}^B$),
we find that the minimum free energy is not 
obtained when $\Delta_{0}$ and $\Delta _{1}$ coexist, but when either
$\Delta_{0}$ or $\Delta_{1}$ vanishes. Then, the total free energy $F_{T}$
reduces to the free energy $F_{0}$ or $F_{1}$ respectively of the 
N=0 and the N=1 FISDW phases given by: 
\begin{eqnarray}
F_{0}-F_{norm}&=&\frac{a_{0}}{2}\left( \Delta_{0}^{A}\right)^{2}+
\frac{a_{0}}{2}\left( \Delta_{0}^B\right)^{2}+\frac{b_{0}}{2}
\left( \Delta_{0}^{A}\right)^{4}\nonumber \\
&+&\frac{b_{0}}{2}\left( \Delta_{0}^{B}\right)
^{4}+c_{0}\left( \Delta _{0}^{A}\right)^{2}\left( \Delta_{0}^{B}\right)^{2}\\
F_1-F_{norm}&=&a_1\left(\Delta_1\right)^{2}+b_1\left(\Delta_1\right)^{4}\nonumber
\end{eqnarray}
The study of the relative stability of the N=0 and the N=1 phases
reduces to the comparison of their corresponding free energies $F_0$ and $F_1$.
%
\begin{figure}[h]
\epsfxsize=5cm
\epsfysize=4cm
\centerline{\epsfbox{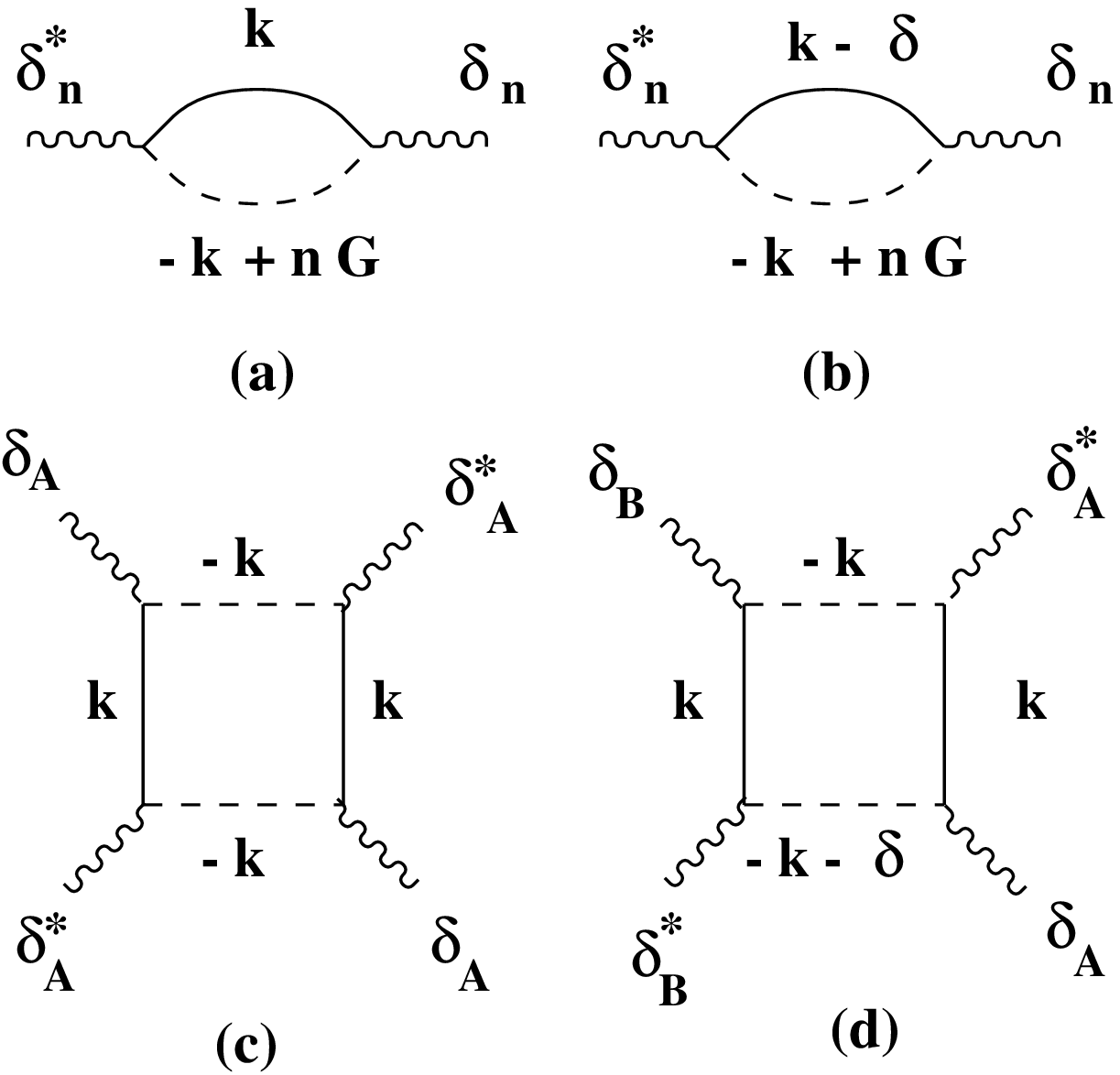}}
{\noindent Fig.2: {\small Diagrammatic representation of the second and fourth order
terms of the free energy in the N=0 state. $\delta=\frac{4\Delta}{v_F}$,
$\delta_m=I_0\Delta_0^m$ (m=A,B) and $\delta_n=I_n\Delta_0^{A,(B)}$.}}
\end{figure}
The second order diagrams (a) and (b) in Fig.2 lead to the 
instability criterion for the N=0 phase, from which we deduce the $T_0$
transition temperature.\newline
The coefficients $b_0$ and $c_0$ of the fourth order terms in eq.3 are
respectively given by (c) and (d) diagrams of Fig.2.
$c_0$ is positive and depends, as $b_0$
on the effective anion gap $\Delta $.

For the N=1 phase, we obtain, as expected, the same Stoner criterion
as in the (TMTSF)$_2$PF$_6$ case\cite{Poil,Maki}.
We denote by $T_1$ the transition temperature, at which the metallic state is 
unstable against the formation of the N=1 SDW state.\newline
Close to $T_{0}$, the N=0 phase is semiconducting rather than insulating, because of
the small value of the order parameter and of the large thermal fluctuations.\newline
We set $\Delta_{0}^{A}=\Delta_{0}^{B}\equiv \Delta_{0}$.
The minimization of $F_0$ and $F_1$, with respect to $\Delta_0$ and $\Delta_1$ shows that 
$\left[ F_{1}\right]_{\min }$ is lowered compared to $\left[ F_{0}\right]_{\min }$ 
for $T<T^{\ast}_1$, where $T_{1}^{\ast}$ satisfies:
$
\left[ F_{1}\right]_{\min}(T^{\ast}_1)=\left[ F_{0}\right]_{\min }(T^{\ast}_1)
$.

By decreasing the temperature, the effect of the coupling term in eq.3 gets more and
more pronounced, because of the enhancement of the order parameters. 
However, the latter are still small enough to justify the Landau expansion at low temperature.
The N=0 phase is then destabilized and the N=1 phase becomes stable. Hence, at
a temperature $T_{1}^{\ast}$, a transition from the N=0
phase to the N=1 phase takes place, with a discontinuity of the order
parameter. Figure 3 shows the field-temperature phase diagram obtained from
Landau calculations. With decreasing temperature, the phase diagram
exhibits the presence of two distinct transitions. First, a second order
transition occurs at $T=T_{0}$ from the metallic state to the N=0 SDW
phase. Then, a first order transition takes place at $T_{1}^{\ast}$ from 
the N=0 phase to the N=1 phase.
According to our calculations $\Delta _{0}$ is smaller
than $\Delta_{1}$ for $T<T_{1}^{\ast}$. We therefore suggest that the first order transition is
a semiconductor-semiconductor transition, in agreement with 
the behavior of the magnetoresistance\cite{Moser,Chung}. 
On the other hand, the Hall conductivity in the inner phase is determined
by $\Delta_{1}$, in accordance with the experimental results,
since, in the case of coexisting order parameters, the Hall conductivity is due the 
largest term \cite{Yakov2}.
%
\begin{figure}[h]
\epsfxsize=5.5cm
\epsfysize=4.5cm
\centerline{\epsfbox{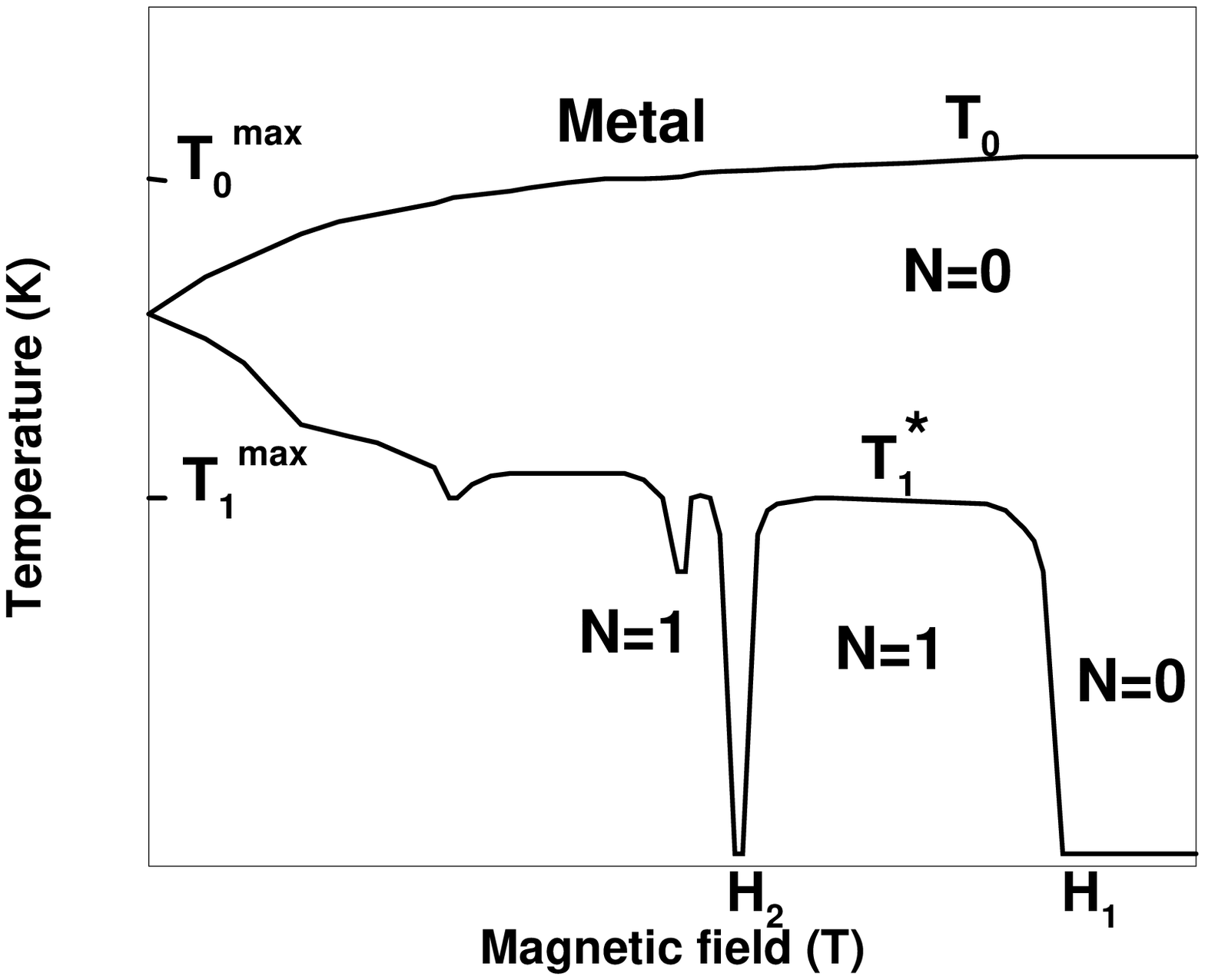}}
{\noindent Fig.3: {\small Temperature-field phase diagram of (TMTSF)$_2$ClO$_4$.
The calculations are done for $E_F=4000$ K, $t_1=250$ K, $t_2=10$ K, $V=8$ K, 
$\frac{g_2}{\pi v_F}\sim 0.13$ for the intraband process and $\frac{g_2}{\pi v_F}\sim 0.17$
for the interband one.
$T_0^{max}\sim 5.6$ K, $T_1^{max}\sim 3$ K, $H_1 \sim 72$ T and 
$H_2 \sim 56$ T.
}} 
\end{figure}
In the very high field regime, the second order transition line persists,
with a nearly field independent critical temperature, while the first order
transition temperature decreases strongly and vanishes at $H=H_1$. 
At this critical field and at low temperature, a transition from the N=1
semiconducting state to the N=0 insulating state takes place.
At lower field the $T_{1}^{\ast}$ line collapses at a critical field $H_2$.
The predicted $H_{1}$ and $H_{2}$ are higher than the experimental values.
However, taking into account the effects of low
dimensional fluctuations by using renormalized Landau parameters might still
improve the quantitative agreement with the experimental data.

It is important to stress a crucial point: the effective coupling constants
$g_2$ are different for intraband and interband nestings, because of different
effects of low dimensional fluctuations \cite{Bourbon1,Bourbon2}. Although the bare couplings are,
of course, the same, our renormalization group calculation, in the presence of
two bands \cite{Kishine} below the anion ordering temperature $T_{AO}$, leads to different
renormalized couplings for intraband and interband processes.
This is an essential feature for the determination of the phase diagram, which,
up to now, has been ignored.

On the other hand, for $H< H_2$, our proposed phase diagram
shows the presence of a new transition line. Chung {\it et al.}
gave recently some arguments in favor of the existence of
such line \cite{Chung}.
Furthermore, we found, as expected \cite{Heritier}, that the decrease 
of the imperfect nesting parameter $t_2$ furthers the formation of N=0 phase.
$H_{1}$ and $H_{2}$ are found to be shifted to lower values. 
The critical temperature $T_{1}^{\ast}$ is decreased, 
whereas $T_0$ is slightly increased.

In summary, we have proposed a new theoretical phase diagram for 
(TMTSF)$_2$ClO$_4$ in the high field regime.
We have discussed the competition between the
high field induced SDW phases. We found that, the N=0 phase has
always the highest critical temperature, but it implies the coexistence of
two different order parameters, corresponding to the two Fermi surfaces.
This coexistence tends to destabilize this phase as the temperature is
lowered, because of the increase of the order parameters.
Eventually, a first order transition occurs, from the N=0 to the N=1 phase.
The proposed phase diagram seems to be in qualitative agreement with the 
experimental data.

%
%
We are grateful to J. Moser and  D. J\'erome for discussing their unpublished data
and we would like to acknowledge N. Dupuis for stimulating discussions.

\end{multicols}
\end{document}